\documentclass[conference,a4paper]{IEEEtran}
\IEEEoverridecommandlockouts

\usepackage[hidelinks]{hyperref}
\usepackage[cmex10]{amsmath}
\usepackage{amssymb,amsfonts}
\usepackage{dblfloatfix}

\usepackage[ruled,vlined]{algorithm2e}
\usepackage{graphicx}
\graphicspath{{Figures/PDF/}{Figures/PNG/}}

\usepackage{booktabs}
\usepackage{siunitx}
\usepackage[numbers,compress]{natbib}
\usepackage{texnames}
\usepackage{bm,bbm}
\usepackage{orcidlink}
\usepackage{titlesec}

\titlespacing\section{0pt}{12pt plus 4pt minus 2pt}{0pt plus 2pt minus 2pt}
\titlespacing\subsection{0pt}{12pt plus 4pt minus 2pt}{0pt plus 2pt minus 2pt}
\titlespacing\subsubsection{0pt}{12pt plus 4pt minus 2pt}{0pt plus 2pt minus 2pt}
\begin{document}
\topmargin=0mm

\title{{Few-shot Human Motion Recognition through Multi-Aspect mmWave FMCW Radar Data\vspace{-0.2em}}
\thanks{$\dag$~Authors contributed equally to this work.

*~Corresponding author. Email address: \href{chenlingfeng@nudt.edu.cn}{chenlingfeng@nudt.edu.cn}.

This work was supported by the China National Natural Science Foundation under Grant 62201588, in part by the Hunan Natural Science Foundation under Grant 2024JJ10007, in part by the National University of Defense Technology Research Program Project under Grant ZK21-14 and ZK23-18, in part by the Project funded by the China Postdoctoral Science Foundation under Grant 2022M723914.
}
}

\author{
\IEEEauthorblockN{Hao Fan$\dag$\orcidlink{0009-0006-8746-0764}}
\IEEEauthorblockA{\textit{College of Electronic Science and Technology}\\
\textit{of National University of Defense Technology}\\
Changsha 410073, China\\
fanhao@nudt.edu.cn} \\
\IEEEauthorblockN{Chengbai Xu$\dag$\orcidlink{0009-0002-5041-786X}}\IEEEauthorblockA{\textit{College of Electronic Science and Technology}\\
\textit{of National University of Defense Technology}\\
Changsha 410073, China\\
xuchengbai@nudt.edu.cn} \\
\IEEEauthorblockN{Jiadong Zhou\orcidlink{0009-0001-7921-7493}}
\IEEEauthorblockA{\textit{College of Electronic Science and Technology}\\
\textit{of National University of Defense Technology}\\
Changsha 410073, China\\
zhoujiadong21@nudt.edu.cn}
\vspace{-3em}
\and
\IEEEauthorblockN{Lingfeng Chen*\orcidlink{0009-0003-2690-9407}}
\IEEEauthorblockA{\textit{College of Electronic Science and Technology}\\
\textit{of National University of Defense Technology}\\
Changsha 410073, China\\
chenlingfeng@nudt.edu.cn} \\
\IEEEauthorblockN{Panhe Hu\orcidlink{0000-0001-5895-3152}}
\IEEEauthorblockA{\textit{College of Electronic Science and Technology}\\
\textit{of National University of Defense Technology}\\
Changsha 410073, China\\
hupanhe13@nudt.edu.cn} \\
\IEEEauthorblockN{Yongpeng Dai\orcidlink{0000-0002-4142-6265}}
\IEEEauthorblockA{\textit{College of Electronic Science and Technology}\\
\textit{of National University of Defense Technology}\\
Changsha 410073, China\\
daiyp3360@126.com
\vspace{-3em}}}
\maketitle
\begin{abstract}
    Radar human motion recognition methods based on deep learning models has been a heated spot of remote sensing in recent years, yet the existing methods are mostly radial-oriented. In practical application, the test data could be multi-aspect and the sample number of each motion could be very limited, causing model overfitting and reduced recognition accuracy. This paper proposed channel-DN4, a multi-aspect few-shot human motion recognition method. First, local descriptors are introduced for a precise classification metric. Moreover, episodic training strategy was adopted to reduce model overfitting. To utilize the invariant sematic information in multi-aspect conditions, we considered channel attention after the embedding network to obtain precise implicit high-dimensional representation of sematic information. We tested the performance of channel-DN4 and methods for comparison on measured mmWave FMCW radar data. The proposed channel-DN4 produced competitive and convincing results, reaching the highest 87.533\% recognition accuracy in 3-way 10-shot condition while other methods suffer from overfitting.
    Codes are available at: \href{https://github.com/MountainChenCad/channel-DN4}{\texttt{https://github.com/MountainChenCad/channel-DN4}}.
\end{abstract}
\vspace{-0.8em}
\begin{IEEEkeywords}
	Few-shot learning, human motion recognition, incomplete target aspect, range-doppler map, channel attention.
\end{IEEEkeywords}
\vspace{-2em}
\section{Introduction}
In recent years, human motion recognition has emerged as a prominent area of research and has been applied in fields including medical monitoring, anti-terrorism \cite{1,2,3}, etc. For existing researches \cite{4,5}, remote sensor-based human motion recognition relies primarily on optical cameras or radar technology (e.g., mmWave FMCW radar). Nevertheless, optical cameras are sensitive to environmental conditions such as lighting and weather, while radar-based techniques offer advantages in this regard and have therefore attracted significant attention from researchers \cite{6,7,8}.

Traditional radar-based motion recognition methods extract information through hand-designed features which can be subjective and can limit their performance in real world scenarios \cite{9, 10, 11,12}. Deep learning-based methods are capable of extracting high-dimensional information, achieving significant improvements in recognition accuracy. Existing works include Convolutional Neural Networks (CNNs) \cite{13,14} and Recurrent Neural Networks (RNNs) \cite{15,16}. Nonetheless, deep models are inherently data-driven, which need a substantial amount of data to achieve optimal performance. The aspect sensitivity of radar data (e.g., Range-Doppler (RD) map) causes significant aspect variations. In non-cooperative scenarios where only a limited number of multi-aspect training data points are available for model training, current deep learning-based methods oriented on radial data can hardly generalize effectively to multi-aspect data. This limitation results in increased model overfitting lead to poor generalization \cite{17}. Hence, enhancing multitask generalization capabilities under few-shot conditions presents an urgent challenge.


To address this challenge, existing methods include data augmentation \cite{18}, model simplification \cite{19} and Few-Shot Learning (FSL) \cite{20, 21, 22, 23, 24, 25}. In the past few years, many influential studies in FSL of optical target recognition offer us inspiration, which aim to enhance model generalization with only a few training samples given, including Model Agnostic Meta-Learning (MAML) \cite{21}, Prototypical Network (ProtoNet) \cite{22}, Deep Nearest Neighbor Neural Network (DN4) \cite{23} and Relation Network (RN) \cite{24}. Yang et al. \cite{25} integrated the Meta-Data-Augmentation (MDA) strategy into the meta-learning process and achieved an impressive result. Gong et al. \cite{8} used a learning technique which put micro-Doppler (m-D) signatures into a designed residual block ProtoNet for training and classification, resulting in significant generalization capability and less computationally complexity, with high recognition accuracy. Nevertheless, current FSL human motion recognition methods mostly apply optical methods, ignoring the unique characteristics of the radar data. This oversight may lead to suboptimal performance due to the domain gap between optical and radar modalities.

\textbf{\textbf{Our Work.}} To achieve the few-shot human motion recognition in multi-aspect conditions, we propose channel-DN4, an improved multi-aspect few-shot human motion recognition method for mmWave FMCW radar. 
1) In response to the problem of network overfitting and poor generalization under the condition of incomplete aspect angle in few-shot RD map data, we proposed channel-DN4, whose local descriptor and episodic training strategy effectively improves the recognition accuracy under incomplete aspect angle conditions. 2) To utilize the invariant information under multi-aspect condition, we added a Squeeze-and-Excitation Network (SENet) after the embedding network, which effectively improved the model performance through channel attention that help extract precise high-dimensional representation of aspect-invariant sematic information. 3) To evaluate the robustness of channel-DN4, experiments on different distance and human target conditions are conducted on measured mmWave FMCW radar data. Codes are available at: \href{https://github.com/MountainChenCad/channel-DN4}{\texttt{https://github.com/MountainChenCad/channel\\-DN4}}.
\vspace{-2em}
\section{Proposed Method}
\subsection{Signal Model}
\begin{figure}[h]
    \centering
    \includegraphics[width=\linewidth]{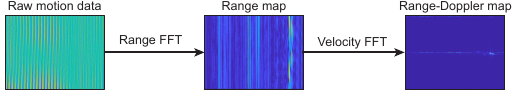}
    \caption{The illustration of processing procedures from radar echo to Range-Doppler map.}
    \label{fig1}
\end{figure}
The flowchart of the processing procedure for this module is shown in Fig. \ref{fig1}.
The radar echo data used in this experiment is taken by frames. Each frame has 256 chirps, and each chirp has 256 sampling points. All chirps emitted within a frame are arranged to obtain the 256×256 raw motion data matrix as shown in the left panel of Fig. \ref{fig1}. Fast Fourier Transformation (FFT) is first performed in the fast time dimension (Range FFT), whose goal is to extract the frequency $\frac{2\mu R}{c}+f_{d}$. Here, $\mu$ represents slope of frequency modulation, $R$ represents the radial distance between the target and the radar, $c$ represents the speed of light, $f_{d}$ represents the Doppler frequency. $\frac{2\mu R}{c}$ is the frequency corresponding to the target distance. Since $f_{d}$ is several orders of magnitude smaller than $\frac{2\mu R}{c}$, it can be ignored, i.e., the Range map shown in the middle of Fig. \ref{fig1} can be obtained by using 1D-FFT on the fast time dimension. Then perform 1D-FFT on the slow time dimension (Velocity FFT), whose goal is to extract the Doppler frequency $f_{d}$, that is, the frequency corresponding to the target velocity. Then the obtained map will go through a normalization process. For the normalization, we define the original two-dimensional RD numerical matrix as $\boldsymbol{X}$. The normalized two-dimensional RD numerical matrix is denoted as $\bar{\boldsymbol{X}}$. The normalization operation is denoted as $
    \bar{\boldsymbol{X}}=(\boldsymbol{X}-\min\left\{\boldsymbol{X}\right\})/\max\left\{\boldsymbol{X}\right\}$
. $\min\left\{\boldsymbol{X}\right\}$ represents the minimum value in the original RD numerical matrix, $\max\left\{\boldsymbol{X}\right\}$ represents the maximum value in the original RD numerical matrix. After normalization, the original RD numerical matrix is mapped to the interval [0,1]. Finally, the normalized RD map shown on the right of Fig. \ref{fig1} is drawn by using the built-in function \texttt{Imagesc()} of Matlab, and the resulting RD map saved as a JPEG file $\hat{\boldsymbol{X}}$. \vspace{-1em}
\subsection{Problem Set-up}
In this section, the general model of the few-shot classification task under multi-aspect conditions is introduced. Since the aim of few-shot learning is to achieve good model performance with only a few training data, current FSL method achieves this objective through S/Q episodic training \cite{26}. We herein introduce the idea of a task $\mathcal{T}$. For each task $\mathcal{T}$, consider RD map sample $\hat{\boldsymbol{X}}_i\in\mathbb{R}^{m\times q}$, where $m,q$ denotes the range and doppler dimension of the RD map respectively. Then a $N$-way $K$-shot task $\mathcal{T}$ of the model can be expressed as
\begin{equation}
\begin{aligned}
    \mathcal{T}=\{\{(\hat{\boldsymbol{X}}_1,l_1),&...,(\hat{\boldsymbol{X}}_{K\cdot N},l_{K\cdot N})\},\{\overline{\hat{\boldsymbol{X}}_1},...,\overline{\hat{\boldsymbol{X}}_t}\};\\&l_i\in\{1,N\}\}
\end{aligned}
\end{equation}
where, $l_i$ is the label of $\hat{\boldsymbol{X}}_i$. $\{\overline{\hat{\boldsymbol{X}}_1},...,\overline{\hat{\boldsymbol{X}}_t}\}$ represents $t$ samples as query set $\mathcal{Q}$ and $\{(\hat{\boldsymbol{X}}_1,l_1),...,(\hat{\boldsymbol{X}}_{K\cdot N},l_{K\cdot N})\}$ stands for $K$-shot support set $\mathcal{S}$. In the experimental sections, we have obtained results when $t=15$ to give a comprehensive evaluation of the methods. $N$ means the number of classes the samples in $\mathcal{T}$ can be divided into. In our multi-aspect scenario, the support $\mathcal{S}$ and the query set $\mathcal{Q}$ comes from different aspects, causing the huge gap of distribution between training and testing data. The output of the classification model can be denoted as $Y=(y_1,\ldots,y_t)\in\{1,N\}$. If we consider our model as $f_\Theta(\mathcal{T})=p(Y|\mathcal{T})$, which is determined by trainable parameters $\Theta$, then our optimization goal of the problem can be expressed as:
\begin{equation}
    \min_{\Theta}\frac{1}{L}\sum_{i\leq L}\nabla\mathcal{L}(f_\Theta(\mathcal{T}_i), Y_i)
\end{equation}
where $\mathcal{L}$ stands for the loss function. $L$ represent the total number of sample tasks $\mathcal{T}_i$, whose corresponding labels are $Y_i$. Cross Entropy (CE) loss is used in this paper for this classification problem, and the CE loss of a task $\mathcal{T}_i$ can be calculated by \eqref{eqce}.
\begin{equation}
\mathcal{L}(f_{\Theta}(\mathcal{T}_i), Y_i)=-\sum_{N}y_N\log{P(Y_{i}=y_N|\mathcal{T}_i})
\label{eqce}
\end{equation} 
\vspace{-2em}
\subsection{Embedding Network}
\begin{figure}[h]
    \centering
    \includegraphics[width=0.75\linewidth]{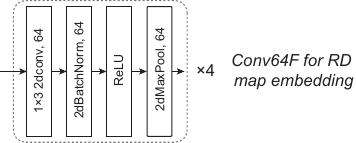}
    \caption{The structure of the Conv64F embedding network.}
    \label{fig4}
\end{figure}

The embedding network employs a simple Conv64F architecture \cite{26} for feature extraction, whose structure is shown in Fig. \ref{fig4}. Let $\mathit{\Psi}_\theta(\cdot)$ represents the embedding network, $\theta$ represents the model parameters which determine the embedding network.
This step can be formulated as $
\mathit{\Psi}_\theta(\hat{\boldsymbol{X}})=\boldsymbol{F}\label{1}
$. The RD map $\hat{\boldsymbol{X}}$ is embedded to obtain the high-dimensional feature vector $\boldsymbol{F} \in \mathbb{R}^{h \times w \times d}$, in which $d$ represents the number of feature channels, and the spatial dimensions are $m=h \times w$, where $h$ for height and $w$ for width.
Following the style of Li et al. \cite{23}, the high-dimensional feature vector $\boldsymbol{F}$ is considered as $m=h \times w$ local descriptors $\boldsymbol{x}_i\in\mathbb{R}^{1\times1\times d}$ in the spatial dimensions, which can be expressed as \eqref{FF}:
\begin{equation}
    \boldsymbol{F}=[\boldsymbol{x}_1,...,\boldsymbol{x}_m]\in\mathbb{R}^{d\times m}\label{FF}
\end{equation}
where the length of each local descriptor is $d$.
These local descriptors improve the model's representation ability of local features compared to a single feature map, enabling the model to precisely compare local similarity in following secessions.
\vspace{-1.5em}
\subsection{Channel Attention}
Compared to Li et al. \cite{23}, the proposed channel-DN4 added a key lightweight channel attention process, SENet, after the embedding network. SENet is typically composed of convolutional layers, SE blocks, and fully connected layers \cite{27}. The SE block is the core module of SENet, which includes a squeeze module and an excitation module, effectively enhancing the network's perception of features and strengthening the model's performance and generalization capabilities in multi-aspect conditions by extracting invariant sematic information.

In the squeeze module, global squeeze pooling is performed on all local descriptors of $\boldsymbol{F}$ in spatial dimensions, which transforms each two-dimensional feature channel into a real number $a_i$. Therefore, to some extent, each $a_i$ has a global receptive field in its corresponding feature channel. Combining each $a_i$ forms the global information vector $\boldsymbol{z}_d=[a_1,...,a_d]\in\mathbb{R}^{1\times1\times d}$, which represents the global distribution of responses across the feature channels. This step can be expressed as \eqref{2}:
\begin{equation}
    \boldsymbol{z}_d=\boldsymbol{F}_{sq}(\boldsymbol{x}_i)=\frac{1}{h\times w}\sum_{j=1}^h\sum_{k=1}^w\boldsymbol{x}_i(j,k)\label{2}
\end{equation}
where $\boldsymbol{F}_{sq}(\cdot)$ represents the squeeze operation.

In the excitation module, the global information $\boldsymbol{z}_d$ obtained from the squeeze module is passed through two fully connected layers and followed by a ReLU activation function layer and a Sigmoid activation function layer to obtain the corresponding channel weight $\boldsymbol{s}\in\mathbb{R}^{1\times1\times d}$. The weights of the fully connected layers are denoted as $\boldsymbol{W}_1,\boldsymbol{W}_2$, and the ReLU activation function layer and Sigmoid activation function layer are denoted as $\mathrm{ReLU}\left(\cdot\right),\sigma\left(\cdot\right)$. A scaling parameter $r$ is set to reduce the number of channels and decrease the computational load in subsequent operations. The excitation operation can be represented as $\boldsymbol{F}_{ex}(\cdot)$, and the specific operations can be expressed as \eqref{3}:
\begin{equation}
    \boldsymbol{s}=\boldsymbol{F}_{ex}(\boldsymbol{z}_d)=\sigma(\boldsymbol{W}_2\mathrm{ReLU}(\boldsymbol{z}_d,\boldsymbol{W}_1))\label{3}
\end{equation}
where $\boldsymbol{W}_1\in\mathbb{R}^{d/r\times d}$, $\boldsymbol{W}_2\in\mathbb{R}^{d\times d/r}$, $\boldsymbol{s}$ characterizes the weights of the $d$ feature maps in feature $\boldsymbol{F}$. Since $\boldsymbol{s}$ is learned through fully connected layers and non-linear layers, it can be trained end-to-end. By multiplying each channel in $\boldsymbol{F}$ with its corresponding weight $\boldsymbol{s}$, a new feature vector $\hat{\boldsymbol{F}}$ is obtained.The specific operations can be expressed as $\hat{\boldsymbol{F}}=\boldsymbol{s}\times\boldsymbol{F}$. Here, $\hat{\boldsymbol{F}}=[\hat{\boldsymbol{x}}_1,...,\hat{\boldsymbol{x}}_m]\in\mathbb{R}^{d\times m}$, and $\hat{\boldsymbol{x}}_i\in\mathbb{R}^{1\times1\times d}$ represents the local descriptors after channel attention.
\vspace{-1.5em}
\subsection{Classification Metric}
For each local descriptor $\hat{\boldsymbol{x}}_i\in\mathbb{R}^{1\times1\times d}$ of $\hat{\boldsymbol{F}}$, we calculate its $k$-nearest neighbors ($k$-NN) sum of Cosine Similarity (CS) for classification. In the $N$-way $K$-shot training and testing process, for each local descriptor $\hat{\boldsymbol{x}}_i$, the $\boldsymbol{K}$ closest local descriptors are found in each class $l_{i}$. This allows for the calculation of the sum of CS with all classes, the class with the maximum a posteriori probability is taken as the recognition result. The above operations can be formulated as \eqref{5} and \eqref{6}:
\begin{equation}
\mathbf{\Phi}\left(\hat{\boldsymbol{F}},l_{i}\right)=\sum_{p=1}^{m}\sum_{q=1}^{k}\cos\left(\hat{\boldsymbol{x}}_{p},\mathring{\boldsymbol{x}}_{p}^{q}\right)\label{5}
\end{equation}
\begin{equation}
\cos(\hat{\boldsymbol{x}}_p,\mathring{\boldsymbol{x}}_p)=\frac{\hat{\boldsymbol{x}}_p^\top\mathring{\boldsymbol{x}}_p}{|\hat{\boldsymbol{x}}_p|\cdot|\mathring{\boldsymbol{x}}_p|}
\label{6}\end{equation}
where $\mathbf{\Phi}(\cdot)$ represents the metric operation of the sum of the nearest neighbour CS, $\cos(\cdot)$ denotes the operation of obtaining the CS, $\mathring{\boldsymbol{x}}_i\in\mathbb{R}^{1\times1\times d}$ represents the local descriptors corresponding to the label, $\left|\cdot\right|$ signifies the modulus operation, and $(\cdot)^\top$ indicates the transpose operation of a vector.
\vspace{-1em}
\section{Experiments}
\subsection{Experimental Settings}
\subsubsection{Data Measurement}
\begin{figure}[t]
    \centering
    \includegraphics[width=\linewidth]{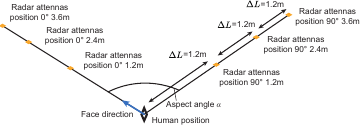}
    \caption{A photo of the experimental scenario and corresponding illustrations.}
    \label{fig2l}
\end{figure}

The specific scenario of our measured dataset construction is shown in the Fig. \ref{fig2l}. Two mmWave FMCW radars are with the following specific parameters: 2GHz radar bandwidth, 78GHz\textasciitilde80GHz operating frequency, 102.4ms chirp period, 2GHz/102.4ms frequency modulation rate, 2.5MHz sampling rate, the number of sampling points per chirp is 256, and the number of chirps per frame is 256. The radar antenna is set 1.2m above the ground. At 0$^\circ$ and 90$^\circ$ radially from the face direction of the volunteer, three sets of positions were set at distances $L$ 1.2m, 2.4m, and 3.6m with intervals $\Delta L$ of 1.2m. Each set was composed of two corresponding positions. During the experiments at each set of positions, two mmWave FMCW radars were placed at the two positions respectively towards the volunteer forming an angle $\alpha$ which we set to be 90$^\circ$. Three volunteers with heights of 1.92m, 1.77m, and 1.70m performed 9 types of motions, including squatting, waving the right hand, oscillating the right hand vertically, circling with both hands, right-hand circling, standing, rotating, left-hand circling, and left-hand vertical oscillation. When collecting data, both radars operate simultaneously, and the collection time for each motion is controlled at 10s. A multi-aspect human motion recognition dataset consisting of 1807 RD maps was obtained (about 200 frames for each type of motion, 100 frames for each aspect).

\begin{table}[t]
	\centering
	\caption{Accuracy with height of 1.92m and distance of 1.2m.}\label{Tab2}
	\begin{tabular}{r c c c}
		\toprule
		\textbf{Methods} & \textbf{3-way 1-shot} & \textbf{3-way 5-shot} & \textbf{3-way 10-shot}\\ \cmidrule(lr){1-1} \cmidrule(lr){2-2}\cmidrule(lr){3-3}\cmidrule(lr){4-4}
		Baseline \cite{26}&63.822&77.467&81.733\\
		Baseline++ \cite{26}&69.156&77.089&80.444\\
		RelationNet \cite{24}&33.333&33.333&33.333\\
        MAML \cite{21}&61.333&70.333&62.311\\
		ProtoNet \cite{22}&54.867&57.200&63.067\\
        DN4 \cite{23}&66.156&81.622&86.089\\
        \bf{Ours}&\bf{69.333}&\bf{83.356}&\bf{87.533}\\\bottomrule
	\end{tabular}
\end{table}
\begin{table}[t]
	\centering
	\caption{Accuracy of different $L$ in 3-way 10-shot, $H$=1.92m condition.}\label{dis}
	\begin{tabular}{r c c c}
		\toprule
		\textbf{Methods} & \textbf{1.2m} & \textbf{2.4m} & \textbf{3.6m}\\ \cmidrule(lr){1-1} \cmidrule(lr){2-2}\cmidrule(lr){3-3}\cmidrule(lr){4-4}
		Baseline \cite{26}&81.733&61.689&70.600\\
		Baseline++ \cite{26}&80.444&73.289&68.400\\
		RelationNet \cite{24}
        &33.333&33.333&33.333\\
        MAML \cite{21}&62.311&33.867&36.200\\
		ProtoNet \cite{22}
        &63.067&61.489&65.356\\
        DN4 \cite{23}
        &86.089&81.222&\bf{80.956}\\
        \bf{Ours}&\bf{87.533}&\bf{82.756}&80.667\\\bottomrule
	\end{tabular}
\end{table}

\begin{table}[t]
	\centering
	\caption{Accuracy of different $H$ in 3-way 10-shot, $L$=1.2m condition.}\label{per}
	\begin{tabular}{r c c c}
		\toprule
		\textbf{Methods} & \textbf{1.92m} & \textbf{1.77m} & \textbf{1.70m}\\ \cmidrule(lr){1-1} \cmidrule(lr){2-2}\cmidrule(lr){3-3}\cmidrule(lr){4-4}
		Baseline \cite{26}&81.733&53.600&71.422\\
		Baseline++ \cite{26}&80.444&74.778&76.444\\
		RelationNet \cite{24}&33.333&33.333&33.333\\
        MAML \cite{21}&62.311&35.044&41.956\\
		ProtoNet \cite{22}&63.067&60.844&66.533\\
        DN4 \cite{23}&86.089&79.333&84.489\\
        \bf{Ours}&\bf{87.533}&\bf{80.067}&\bf{84.778}\\\bottomrule
	\end{tabular}
\end{table}
\vspace{-1em}
\subsubsection{Experimental Platform}
The experiments were conducted on a 64-bit Linux operating system, configured with an Intel(R) Core™ Ultra 9 185H @2.30GHz processor, 32GB RAM, an NVIDIA GeForce RTX 4050 Laptop GPU with 6GB memory. The code was developed using the PyTorch 2.2.0 framework, and the runtime environment was based on CUDA 12.5 and Python 3.8. The measured dataset was splitted to the training and testing datasets $D_{tra}, D_{tes}$. The motions obtained from 0$^\circ$ forms $D_{tra}$, while the rests from 90$^{\circ}$ forms $D_{tes}$. The training epoch was set to 100, with 15 RD maps forming each query set ($t=15$) used for both training and testing. Batch Size is 64. The training dataset was employed to conduct 9-way 5-shot training on the network. During testing, the performance under conditions of 3-way 1-shot, 3-way 5-shot, and 3-way 10-shot scenarios was evaluated respectively. For $k$-NN, we took $k=3$, following Li et al. \cite{23}.
\vspace{-1em}
\subsection{Comparative Experiments}
To compare the performance between the proposed channel-DN4 and classic few-shot classification methods, comparative experiments were conducted on various methods, including the mainstream FSL methods Baseline \cite{26}, Baseline++ \cite{26}, MAML \cite{21}, ProtoNet \cite{22}, RelationNet \cite{24}, DN4 \cite{23} and channel-DN4. The parameters of each method are 0.22M, 0.12M, 0.23M, 0.13M, 0.11M, 0.11M and 0.11M repectively. 
Results shown in TABLE \ref{Tab2} indicate that compared to MAML, ProtoNet, and RelationNet, which do not employ local descriptors, DN4 and channel-DN4 exhibit significantly better recognition performance. Compared to the simple Baseline and the improved Baseline++, channel-DN4 demonstrates superior learning performance, with experimental outcomes aligning with theoretical expectations. The ablation of our method can be observed in the comparison of the channel-DN4 and DN4. Compared to DN4, channel-DN4 utilizes SENet for channel attention, and its performance is on average better than DN4 under 1-shot, 5-shot, and 10-shot conditions by approximately 2.118\%. In terms of the size of the parameter volume, channel-DN4 is equivalent or lower than most methods with Conv64F.
\vspace{-2.5em}
\subsection{Multi-Scenario Study}
To explore the robustness of this method in human motion recognition at different distances and with different human targets (different volunteers), experiments were conducted with three types of distances $L$ and three different human targets of different heights $H$. The results are shown in the TABLE \ref{dis} and TABLE \ref{per}. From TABLE \ref{dis} we can infer that as $L$ increases, the recognition accuracy generally tends to decline. However, channel-DN4 maintained an accuracy above 80\%, outperforming other methods. From TABLE \ref{per}, it can be observed that when the target human subjects vary, the proposed channel-DN4 demonstrates arguable competitive robustness, with an average accuracy rate approximately 0.822\% higher than DN4 and also higher than other methods. A possible explanation for the comparatively lower accuracy of the volunteer with 1.77m can be that during the experiment, there was no regulation on the magnitude of the motion, and the volunteer with 1.77m had relatively smaller motion amplitudes. Overall, compared to the current mainstream few-shot classification methods, channel-DN4 shows better robustness and recognition performance with different distances and human targets.
\vspace{-1em}
\section{Conclusion}
This paper proposed channel-DN4, a innovative design that combines using local descriptors to extract precise local feature representation with channel attention for invariant sematic information in multi-aspect conditions, while adapting episodic training strategy to reduce overfitting. Experimental results on measured mmWave FMCW measured dataset show that channel-DN4 outperforms most of the current mainstream methods in terms of accuracy and robustness. However, the dataset constructed in our work only includes perspectives of 0$^\circ$ and 90$^\circ$. Additionally, the performance of RD diagrams compared to other types of inputs as well as multi-feature fusion remains a promising direction for future work.
\bibliographystyle{IEEEtranN}
\bibliography{references}

\end{document}